# Boron Nitride Nanotube Films Grown From Boron Ink Painting

Lu Hua Li,[1,2]* Ying Chen,[1]* Alexey M. Glushenkov[1]

[1]Institute for Technology Research and Innovation, Deakin University, Waurn Ponds, VIC 3217, Australia
[2]Department of Electronic Materials Engineering, Research School of Physics and Engineering, The Australian National University, Canberra, ACT 0200, Australia

luhua.li@deakin.edu.au; ian.chen@deakin.edu.au

*The growth of nanotube films on various substrates and surfaces is vital for applications in nanoscale functional devices. We report a simple and versatile boron (B) ink painting method that enables high-density boron nitride nanotube (BNNT) films with any desired pattern to be grown on, and firmly attached to, different surfaces. In the method, special B ink is first painted, sprayed or inkjet printed at desired location with required pattern, and then the ink layer is annealed in a nitrogen-containing atmosphere to form BNNT film. The B ink is a liquid mixture of ball-milled B particles, metal nitrate and ethanol. This is the first method capable of growing BNNTs on complex non-flat surfaces, which greatly broadens the potential application of BNNTs. For example, it is demonstrated that a BNNT coated steel mesh can separate water and oil on a microliter scale; a needle given an internal BNNT coating could greatly enhance microfluidic transport; and a coated screw could be used to minimize wear at the interface.*

## INTRODUCTION

Boron nitride nanotubes (BNNTs) have many appealing properties and applications. BNNTs have a rather uniform wide bandgap,[1] and their special optical properties may find various applications in optoelectronics.[2,3] The strong resistance to oxidation makes them more reliable in high temperature environments than carbon nanotubes (CNTs).[4] The piezoelectricity of BNNTs is ideal for building nanoscale electromechanical devices.[5,6] BNNTs enriched in isotopic $^{10}$B provide multifunctional radiation shielding[7] and have excellent thermal conductivity.[8] BNNTs were first synthesized by Zettl's group in 1995.[9] In the past decade, BNNTs in loose powder form have been produced by several groups,[9-14] but there are only a few reports on the growth of BNNT films on flat substrates.[15,16] It has been demonstrated for CNTs that the growth of films on various substrates and surfaces is vital for applications in nanoscale functional devices.[17,18]

Here, we report a unique and versatile boron (B) ink painting method that enables high-density BNNT films with any desired pattern to be grown on, and firmly attached to, different surfaces. A special B ink, a mixture of nanosized B particles, metal nitrate catalyst and ethanol, is first painted, sprayed, or inkjet printed on substrate surface, depending on the desired result, and then annealed in a nitrogen containing gas to convert into a BNNT film. Importantly, this is the first method capable of growing BNNTs on complex non-flat surfaces and greatly broadens the potential application of BNNTs. For example, it is demonstrated that a BNNT covered steel mesh can separate water and oil on a microliter scale; a needle given an internal BNNT coating could greatly enhance microfluidic transport; and a coated screw could be used to minimize wear at the interface.

## EXPERIMENTAL

**B ink preparation.** The ball-milling treatment was conducted in a vertical rotating high-energy ball mill. Several grams of amorphous B powder (95%-97%, Fluka) were sealed in a steel milling vial with four hardened steel balls in anhydrate ammonia (NH$_3$) at 300 kPa. The milling lasted for 100 h at ambient temperature. The milled samples were then mixed with nitrate in ethanol inside a gloved box filled with an inert gas, using an ultrasonic bath for 1 h. The painted substrates were heated in a horizontal tube furnace at 950-1300 °C in N$_2$+15%H$_2$ or NH$_3$ gas. The SiO$_2$ coated Si (SiO$_2$/Si) substrates were produced by annealing the (400) Si wafer in high purity oxygen (O$_2$) at 1100 °C for 3 h.

**Inkjet-printing.** A commercial Canon BJC 210 inkjet printer (cartridge models BC-02 or BX-02) was used for inkjet printing. B ink was poured into the ink chamber of an empty cartridge. Small SiO$_2$/Si wafer pieces (20 mm x 10 mm) were glued on A4 paper for printing. The "nano" patterns were created in Word software. After printing, the SiO$_2$/Si pieces were peeled off the paper and placed in the tube furnace for annealing at 1100 °C in N$_2$+15%H$_2$ for 0.5 h.

**Adhesion test by DMA.** A Q800 DMA (TA Instruments, USA) equipped with a standard tension clamp was used for the adhesion test. The resolution of the loaded force is 0.00001 N. Double-sided Scotch tape (1 cm x 1 cm) was attached to a relatively thick BNNT film (to avoid the tape directly contacting the substrate) and the other side of the tape was fixed to a thin rectangular stainless steel sheet (thickness 0.14 mm). The coated substrate and the stainless steel sheet were fixed to the tension clamps. A shear force (in controlled force mode) was applied to pull the tape away from the BNNT film (Figure 4d). The starting force and the rate of force increase were 0.001 N and 0.5 N/min, respectively.

**Non-flat surface coating.** The woven steel mesh was thoroughly brushed with the B ink (83.3 mg/mL of B in 0.03 M Fe(NO$_3$)$_3$ ethanol). The brushed mesh was hung over an alumina boat and immediately annealed at 1100 °C for 1 h in N$_2$+15%H$_2$ gas, so that both sides of the mesh could contact with the reaction gas and thus be totally covered by BNNTs. The steel needle, cut from a syringe, was totally immersed in B ink (62.5 mg/mL of B in 0.04 M Co(NO$_3$)$_2$ ethanol) so that the ink also covered the needle's internal channel. The ink was brushed on the screw.

**Contact angle.** The contact angles were measured using an optical contact angle and surface tension instrument (CAM 101, KSV) in a conditioned laboratory (20±2 °C, 65±2% RH). The contact angles and drop volumes were automatically calculated by software which uses the Young-Laplace equation to curve-fit drop profiles captured by a camera. The images showing paraffin oil drops (LabChem, density 0.87 g/cm$^3$) passing through the BNNT coated mesh were recorded at a frame interval of 33 ms.

**Other characterization techniques.** Hitachi 4300SE/N and Hitachi 4500 FESEM scanning electronic microscopes at 3 kV were employed to examine sample morphology, and EDS was conducted on the 4300SE/N instrument at 15 kV. TEM investigations were performed using a Philips CM300 (300 kV) microscope.

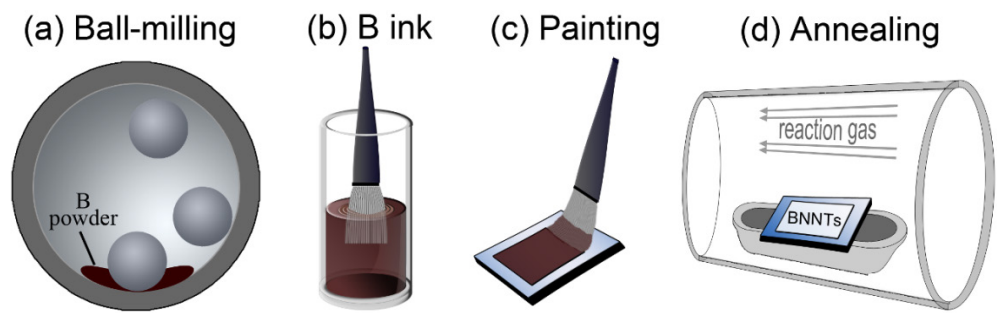

*Figure 1.* Schematic diagram showing the four steps involved in the B ink painting method. a) Ball-milling of B powder in NH$_3$ to produce nanosized B particles. b) Mixing ball-milled B particles, metal nitride and ethanol to form B ink. c) Painting the B ink on the substrate. d) Annealing of the painted substrate in a nitrogen-containing atmosphere to grow the BNNT film.

**RESULTS AND DISCUSSION**

The B ink method, as illustrated in Figure 1, consists of four steps. (a) Nanosized and chemically active B particles are produced by ball-milling of amorphous B powder.[10,19] (b) The ball-milled B particles are homogenously mixed with ethanol and a small amount of ferric (or any other metal) nitrate to form an ink-like solution (B ink) with an ultrasonic bath treatment. (c) The B ink is painted, sprayed or inkjet printed to create a thin layer of B ink in desired pattern on the substrate. (d) The B ink covered substrate is annealed at 950-1300 °C in $N_2+15\%H_2$ or $NH_3$ gas which converts the B layer into a high purity BNNT film.

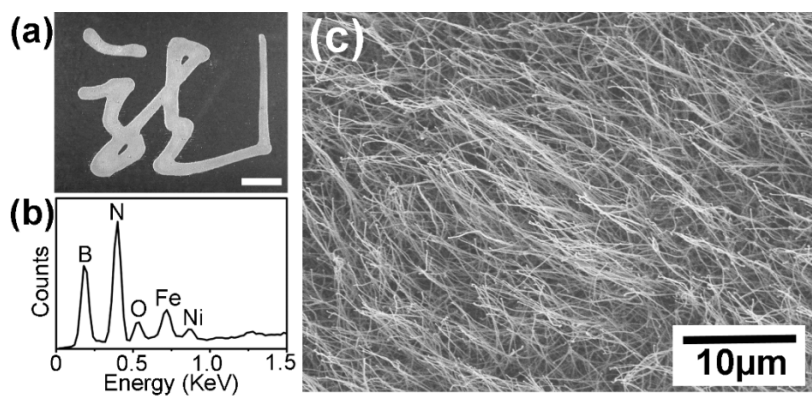

*Figure 2.* Patterned BNNT film produced by B ink painting. a) Optical microscope photo of a white BNNT film with a complicated calligraphic pattern, the scale bar is 5 mm; b) EDS spectra from the BNNT layer; c) SEM image showing high-density BNNTs in the film.

Figure 2a shows an optical microscope image of a patterned white BNNT film formed on a stainless steel substrate by painting the B ink (62.5 mg/mL of B particles in 0.04 M $Fe(NO_3)_3$ ethanol) with a brush and annealing at 1100 °C for 30 min in $N_2+15\%H_2$ atmosphere. A complicated pattern (a traditional Chinese calligraphic dragon character) was chosen to demonstrate the flexibility and ease of use of the method. Regular BNNT patterns could also be produced by mask coating, in which B ink is brushed on the un-covered area of the mask before annealing. The chemical composition of the BNNT films was determined by x-ray energy dispersive spectroscopy (EDS) attached to a scanning electron microscope (SEM) (Figure 2b). Fe and Ni came from both the steel substrate and metal catalysts in the film. The SEM image in Figure 2c shows a high density of BNNTs with diameters of 40-80 nm.

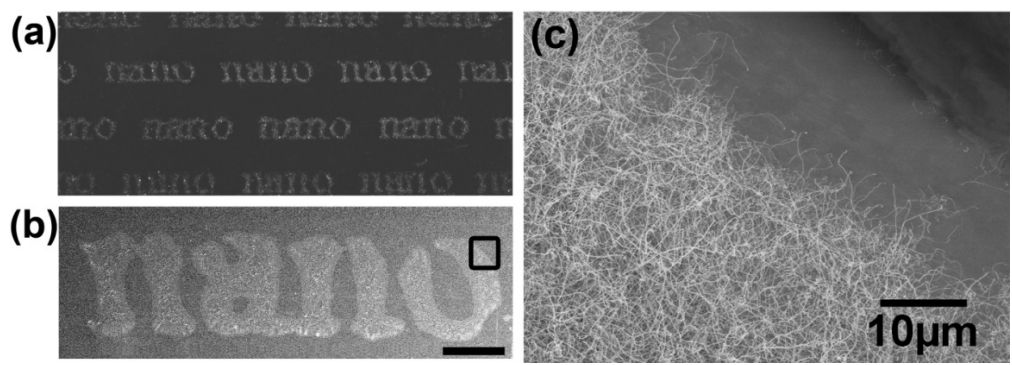

*Figure 3.* BNNT films produced by B ink printing. a) Optical microscope photo of inkjet printed arrays of BNNTs forming the word "nano"; b) SEM image of one word in "Times New Roman" font, scale bar 500 μm; c) higher magnification SEM image of the top right of the letter "o", indicated by the square.

Because of the fineness of the B particles (average size 45 nm), B ink is compatible with inkjet printing. Figure 3a shows an optical microscope photo of BNNTs forming repeated lines of the word "nano" on a SiO$_2$ coated Si (SiO$_2$/Si) substrate. A commercial office inkjet printer with a dilute B ink (50 mg/mL of B in 0.035 M Co(NO$_3$)$_2$ ethanol) were used, and the annealing was conducted at 1100 °C for 30 min in N$_2$+15%H$_2$. Figure 3b shows a SEM image of one group of letters in 'Times New Roman' font of font size "5". A higher magnification SEM image from the edge of the letter "o" (indicated by the square) shows that only the ink printed area has BNNTs (Figure 3c). By using a high resolution jet printing device,[20] the B ink printing method should be able to replicate any BNNT pattern with micron resolution.

In the ink, the ball-milled B particles with metastable and chemically active structures can react with the nitrogen-containing gas at relatively lower temperature;[10,19] the added nitrate provides adequate catalyst for high density nanotube growth; and the ethanol not only makes the growth of BNNT film possible but also slightly enhances the nitriding reactions during annealing.[21] As a result, the high-density BNNT films can be achieved. The BNNT film thickness (nanotube length and density) can be controlled by the amount of ink, that is, the thickness of the ink layer painted on the substrate. Painting or brushing usually resulted in films with a thickness of 5-40 μm. Spraying a diluted B ink (e.g. 30 mg/mL of B in 0.02 M Fe(NO$_3$)$_3$ ethanol) produced a sparse array of BNNTs. The BNNTs produced in a N$_2$+15%H$_2$ atmosphere at 1100 °C, usually with diameter 40-80 nm, can have either a cylindrical or bamboo-like structure. Thin cylindrical BNNTs with diameters less than 10 nm can be produced by annealing in NH$_3$ at 1300 °C.[7] The growth mechanisms of different BNNT structures in different atmospheres have been discussed previously.[22]

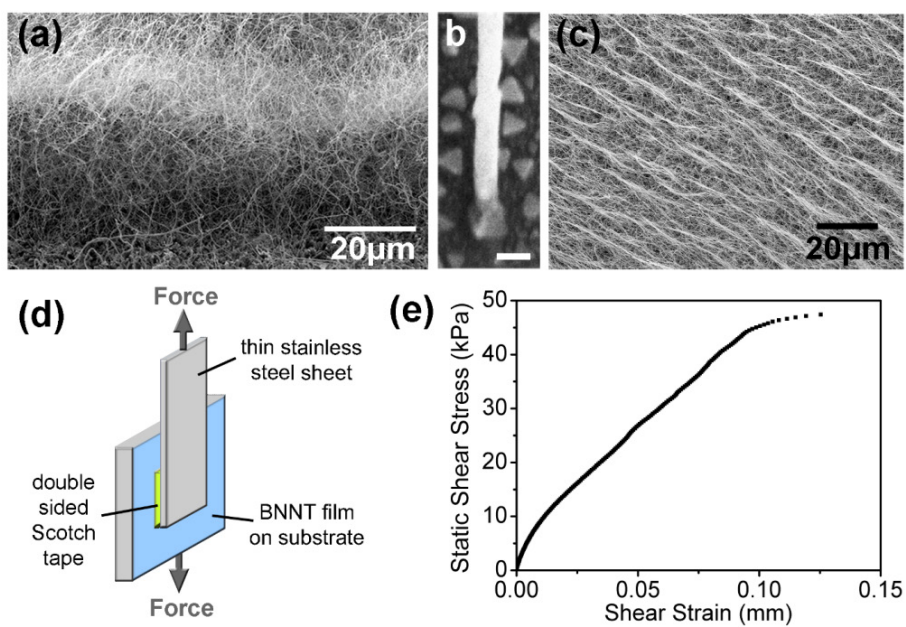

*Figure 4.* a) Cross-sectional view of partly erect BNNTs in the film; b) a BNNT securely attached to the steel substrate via a catalyst particle; c) SEM image of BNNT film after high-pressure air blowing at a short distance; d) schematic diagram showing the set-up of the adhesion measurement; e) the stress-strain curve of a 1 cm$^2$ BNNT film under controlled shear force.

The grown BNNT films are firmly attached to the substrate. BNNTs usually have one end attached to the surface (Figure 4a), because the coating involves nanotube growth during annealing. Some tubes appear to be attached via a metal particle, as shown in the SEM image of Figure 4b. The attachment strength between BNNTs and the substrate was firstly tested by blowing the film with an air gun at a short distance. After high-pressure air blowing, most nanotubes remained, and the strong air flow aligned the tubes to the air flow direction (Figure 4c). The adhesion strength between a 1 cm$^2$ area of the BNNT film and the steel substrate was also measured using a dynamic mechanical analyser (DMA).[23,24] Figure 4d shows the set-up of the measurement. A typical stress-strain curve is shown in Figure 4e, in which the Scotch tape was completely detached from the BNNT film at a shear stress of 4.75 kPa. The average failure shear stress from five BNNT films (all on steel substrates) was 4.67±0.35 kPa. After the tape was pulled off, there was still a white layer of BNNTs left on the substrate, which indicates that the tape did not have direct contact with the substrate and the tape could only pull off some of the BNNTs in the films.

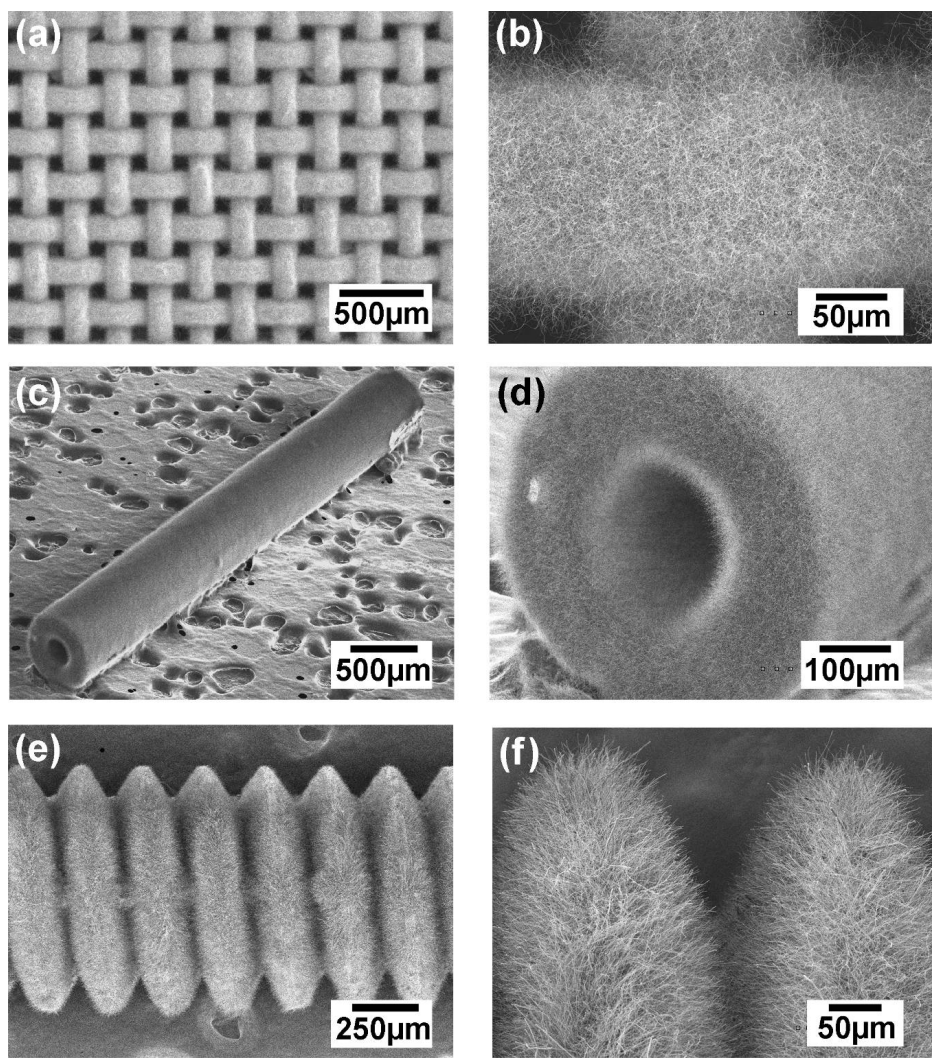

*Figure 5.* SEM images of BNNTs grown on the surfaces of irregularly-shaped objects: a), b) woven steel mesh of wires of 150 μm diameter; c), d) steel syringe needle with outer diameter 510 μm and inner diameter 260 μm; e), f) steel screw of 0.9 mm diameter and the thread spacing of 200 μm.

Another important advantage of the B ink method is that it can grow BNNTs on objects of various and complex shapes. SEM images in Figure 5a and 5b show a steel mesh homogeneously coated with BNNTs. The mesh was thoroughly brushed with the B ink and, after annealing at 1100 °C for 1 h in $N_2+15\%H_2$ gas, both sides were covered by a high density of BNNTs without blocking the small square openings. Figure 5c and 5d show BNNTs were grown on both the external and internal surfaces of a steel syringe needle. This was accomplished by first immersing the needle in the ink and then placing it along the gas flow direction during annealing so that erect BNNTs even grew on the surface of its internal channel. Figure 5e and 5f show BNNTs standing on both the threads and the gaps of a tiny screw. These examples indicate that BNNTs can be grown on any irregularly shaped object, as long as the B ink can reach and cover its surface.

Coating of BNNTs on non-flat objects opens up many new applications. A BNNT covered needle (Figure 5d) could be very useful in microfluidic devices.[25] Figure 6a shows an almost spherical water drop on a BNNT film coated on a flat steel substrate with contact angle 172.1° and hysteresis several degrees.[26,27] This superhydrophobicity is mainly due to the huge surface roughness of the BNNT film that enables the Cassie state.[28] Because of the superhydrophobicity, when fluid flows through such a BNNT coated needle, the adsorption onto the channel surface and the frictional drag can be greatly reduced, allowing better performing microfluidic devices. In contrast to the super anti-wetting to water, the BNNT films are extremely wettable to oil. In general, if a solid has surface tension larger than a quarter of the surface tension of an oil, the solid is wettable to the oil.[29] The surface tension of an individual BNNT with diameter of 40 nm has been measured to be 26.7-27.0 mN/m,[30] and most oils have surface tensions of 20-30 mN/m.[29] So BNNTs are inherently wettable to oil. In addition, the huge roughness of the BNNT film can further increase this wettability, according to the Wenzel equation.[31] As a result, the contact angle of the BNNT film to paraffin oil was found to be ~5° (measured at 1 s after drop dispense). The combined superhydrophobicity and superoleophilicity of the BNNT coated mesh (Figure 5b) opens up exciting possibilities such as the separation of water and oil in microliter volumes.[32] Figure 6b shows that the coated mesh is impermeable to water so that a water drop sits on it with contact angle of 161°. While Figure 6c shows that a paraffin oil drop can pass through the same mesh in just 0.3 second. A third possible application area is low friction coating. A BNNT film on a screw (Figure 5f) has the potential to act as a solid lubricant,[33,34] which minimizes wear even in vacuum or at high temperature, because of the excellent tribological properties of hBN materials and the strong adhesion of the BNNTs to the object surface.

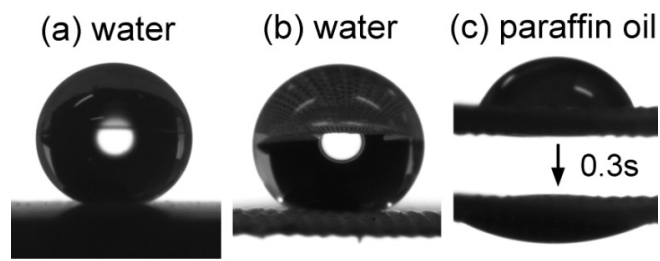

*Figure 6.* (a) A nearly spherical water drop (3.5 μL) on a BNNT film; (b) a water drop (7.2 μL) on the BNNT coated mesh; (c) a paraffin oil drop passing through the BNNT coated mesh in 0.3 s.

**CONCLUSION**

A unique and versatile boron ink method has been developed to effectively grow BNNTs on different surfaces. The innovation of this method is that a high-density and firmly attached BNNT film can be easily accomplished by first painting, mask coating, spraying or inkjet printing the B ink at desired location with required pattern, and then thermally annealing the ink layer in a nitrogen-containing atmosphere. The B ink is prepared by mixing ball-milled B particles with metal nitrate in ethanol. Because of the large capacity of the ball-milling facility, the preparation of B ink can be easily scaled up. No pre-deposition of catalyst is required, as the ink already contains catalyst. The size and nanostructure of BNNTs can be fully-controlled in the thermal annealing process. For the first time, BNNTs can be grown on the surface of objects with irregular or complicated shapes, which opens up many new applications. As examples, a needle with BNNT coating inside the channel could enable better performing microfluidic devices; a BNNT covered mesh can separate oil from water on a microliter scale; and a BNNT film on a screw gives a possible way to minimize wear and solve interfacial problems.

*ACKNOWLEDGMENT -* Authors thank the staff from the Electron Microscopy Unit, Australian National University and Mr. Rob Pow of Deakin University for their valuable assistance with various analyses. We are grateful to Dr. Peter Lamb for assistance in preparing the manuscript. Financial support from the Australian Research Council under the Centre of Excellence program is also acknowledged.

## REFERENCES

[1] J. Yu, D. Yu, Y. Chen, H. Chen, M.-Y. Lin, B.-M. Cheng, J. Li, W. Duan, *Chem. Phys. Lett.* **2009**, 476, 240.

[2] J. S. Lauret, R. Arenal, F. Ducastelle, A. Loiseau, M. Cau, B. Attal-Tretout, E. Rosencher, L. Goux-Capes, *Phys. Rev. Lett.* **2005**, 94, 037405.

[3] C. H. Park, C. D. Spataru, S. G. Louie, *Phys. Rev. Lett.* **2006**, 96, 126105.

[4] Y. Chen, J. Zou, S. J. Campbell, G. Le Caer, *Appl Phys Lett* **2004**, 84, 2430.

[5] E. J. Mele, P. Kral, *Phys. Rev. Lett.* **2002**, 88, 056803.

[6] P. J. Michalski, N. Sai, E. J. Mele, *Phys. Rev. Lett.* **2005**, 95, 116803.

[7] J. Yu, Y. Chen, R. G. Elliman, M. Petravic, *Adv Mater* **2006**, 18, 2157.

[8] C. W. Chang, A. M. Fennimore, A. Afanasiev, D. Okawa, T. Ikuno, H. Garcia, D. Y. Li, A. Majumdar, A. Zettl, *Phys. Rev. Lett.* **2006**, 97, 085901.

[9] N. G. Chopra, R. J. Luyken, K. Cherrey, V. H. Crespi, M. L. Cohen, S. G. Louie, A. Zettl, *Science* **1995**, 269, 966.

[10] Y. Chen, J. D. Fitz Gerald, J. S. Williams, S. Bulcock, *Chem. Phys. Lett.* **1999**, 299, 260.

[11] D. Golberg, Y. Bando, *Appl Phys Lett* **2001**, 79, 415.

[12] R. S. Lee, J. Gavillet, M. L. de la Chapelle, A. Loiseau, J. L. Cochon, D. Pigache, J. Thibault, F. Willaime, *Phys. Rev. B* **2001**, 6412, 121405

[13] C. Tang, Y. Bando, T. Sato, K. Kurashima, *Chem. Commun.* **2002**, 1290.

[14] M. W. Smith, K. C. Jordan, C. Park, J. W. Kim, P. T. Lillehei, R. Crooks, J. S. Harrison, *Nanotechnology* **2009**, 20, 505604.

[15] J. S. Wang, V. K. Kayastha, Y. K. Yap, Z. Y. Fan, J. G. Lu, Z. W. Pan, I. N. Ivanov, A. A. Puretzky, D. B. Geohegan, *Nano Lett.* **2005**, 5, 2528.

[16] L. Guo, R. N. Singh, *Physica E* **2009**, 41, 448.

[17] P. M. Ajayan, O. Z. Zhou, *Applications of Carbon Nanotubes*, Springer, Berlin 2001.

[18] S. Iijima, *Physica B* **2002**, 323, 1.

[19] Y. Chen, L. T. Chadderton, J. FitzGerald, J. S. Williams, *Appl Phys Lett* **1999**, 74, 2960.

[20] J. U. Park, M. Hardy, S. J. Kang, K. Barton, K. Adair, D. K. Mukhopadhyay, C. Y. Lee, M. S. Strano, A. G. Alleyne, J. G. Georgiadis, P. M. Ferreira, J. A. Rogers, *Nat. Mater.* **2007**, 6, 782.

[21] L. H. Li, Y. Chen, A. M. Glushenkov, *Nanotechnology* **2010**, 21, 105601.

[22] J. Yu, B. C. P. Li, J. Zou, Y. Chen, *J. Mater. Sci.* **2007**, 42, 4025.

[23] A. Y. Cao, V. P. Veedu, X. S. Li, Z. L. Yao, M. N. Ghasemi-Nejhad, P. M. Ajayan, *Nat. Mater.* **2005**, 4, 540.

[24] L. T. Qu, L. M. Dai, M. Stone, Z. H. Xia, Z. L. Wang, *Science* **2008**, 322, 238.

[25] C. Choong, W. I. Milne, K. B. K. Teo, *Int. J. Mater. Form.* **2008**, 1, 117.

[26] L. H. Li, Y. Chen, *Langmuir* **2010**, 26, 5135.

[27] C. H. Lee, J. Drelich, Y. K. Yap, *Langmuir* **2009**, 25, 4853.

[28] R. Blossey, *Nat. Mater.* **2003**, 2, 301.

[29] K. Tsujii, T. Yamamoto, T. Onda, S. Shibuichi, *Angew. Chem. Int. Ed.* **1997**, 36, 1011.

[30] K. Yum, M. F. Yu, *Nano Lett.* **2006**, 6, 329.

[31] A. Lafuma, D. Quere, *Nat. Mater.* **2003**, 2, 457.

[32] L. Feng, Z. Y. Zhang, Z. H. Mai, Y. M. Ma, B. Q. Liu, L. Jiang, D. B. Zhu, *Angew. Chem. Int. Ed.* **2004**, 43, 2012.

[33] J. J. Hu, S. H. Jo, Z. F. Ren, A. A. Voevodin, J. S. Zabinski, *Tribol. Lett.* **2005**, 19, 119.

[34] G. Yamamoto, T. Hashida, K. Adachi, T. Takagj, *J. Nanosci. Nanotechnol.* **2008**, 8, 2665.